\documentclass[12pt,preprint]{aastex}

\usepackage{epstopdf}
\usepackage{graphicx}
\newcommand\hour{\mbox{$^{\mathrm h}$}}%
\newcommand\minute{\mbox{$^{\mathrm m}$}}%
\slugcomment{Draft as of \today.}

\shorttitle{A FIR Test of Radiative Grain Alignment}
\shortauthors{Vaillancourt \& Andersson}

\begin{document}

\title{A Far-Infrared Observational Test of the Directional
  Dependence\\in Radiative Grain Alignment}

\author{John E. Vaillancourt and B-G Andersson}
\email{jvaillancourt@sofia.usra.edu; bg@sofia.usra.edu}
\affil{SOFIA Science Center, Universities Space Research Association,
  NASA Ames Research Center, Moffett Field, CA}

\begin{abstract}
  The alignment of interstellar dust grains with magnetic fields
  provides a key method for measuring the strength and
  morphology of the fields.  In turn, this provides a means to study
  the role of magnetic fields from diffuse gas to dense star-forming
  regions.  The physical mechanism for aligning the grains has been a
  long-term subject of study and debate.  The theory of radiative
  torques, in which an anisotropic radiation field imparts sufficient
  torques to align the grains while simultaneously spinning them
  to high rotational velocities, has passed a number of
  observational tests.  Here we use archival polarization data in
  dense regions of the Orion molecular cloud (OMC-1) at
  100, 350, and $850\,\micron$ to test the prediction that the
  alignment efficiency is dependent upon the relative orientations of
  the magnetic field and radiation anisotropy. We find that the
  expected polarization signal, with a 180-degree period, exists at
  all wavelengths out to radii of 1.5 arcminutes centered on the
  BNKL object in OMC-1. The probabilities that these signals would occur due to
  random noise are low ($\lesssim$1\%), and are lowest towards BNKL
  compared to the rest of the cloud.  Additionally, the relative
  magnetic field to radiation anisotropy directions accord with
  theoretical predictions in that they agree to better
  than $15\arcdeg$ at $100\,\micron$ and $4\arcdeg$ at $350\,\micron$.
\end{abstract}


\keywords{dust, extinction --- ISM: individual objects (OMC-1) ---
  ISM: magnetic fields --- polarization --- submillimeter: ISM}


\section{Introduction}

Interstellar polarization at optical through millimeter wavelengths
arises from the extinction and emission of light from asymmetric dust
grains aligned with respect to the interstellar magnetic field
\citep[e.g.,][]{hiltner1949b,Hildebrand1988,Anderssonetal2015}.  More
than 60 years after the discovery of the effect in optical
data \citep{Hall1949,Hiltner1949a}, theory and
observations are converging on an empirically
supported explanation of the alignment mechanism.
\citet{DolginovMytrophanov1976} first proposed the radiative
alignment mechanism in which the scattering of light 
results in net torques on the grains.  Detailed scattering
calculations for some specific grain shapes confirmed that the
strength and direction of these torques were sufficient to align
grains with angular momentum vectors parallel to the magnetic field
\citep{DraineWeingartner1997}, as required by observations. The
analytical approximation of \citet{LazarianHoang2007a} showed
excellent agreement with the scattering calculations,
thereby opening a large parameter space where the
problem of grain alignment is probed with high efficiency
\citep[e.g.,][]{LazarianHoang2007a,HoangLazarian2008}.

In the model of radiative alignment torque (RAT) an irregular grain aligns
with the magnetic field if it is paramagnetic and exposed to an
anisotropic radiation field (see reviews by \citealt{Anderssonetal2015}, \citealt{andersson2015}, and
\citealt{Lazarianetal2015}).
Differential scattering of the left- and right-hand circularly
polarized components of the incident radiation transfers angular
momentum from the radiation field to the grain.  For a paramagnetic
grain, the rapid rotation results in a bulk magnetization via the
Barnett effect \citep{Purcell1979}, in which rotational energy is
traded for spin-flips of atomic nuclei in the material. As the grains
Larmor precess about the magnetic field, continued radiative torques
(averaged over the precession period) change the angular momentum
vector orientation of the grain until it reaches an equilibrium state
with the spin parallel to the magnetic field.

The RAT theory makes several predictions,
 including
 that the grain has a net helicity, and is
paramagnetic, 
and the alignment efficiency increases with the radiation intensity.
These lead to a number of observationally testable predictions,
including the unique prediction that
the alignment efficiency depends on the angle between the directions
of the radiation field anisotropy and the magnetic field.  We test
this prediction using the measured polarization fraction to trace the
grain alignment efficiency and the associated polarization position
angle to infer the magnetic field direction, with respect to the
location of known stars.  \citet{AnderssonPotter2010} and
\citet{Anderssonetal2011} used similar comparisons to confirm the
prediction using optical polarization around the star
HD\,97300 for relatively low opacity material.  Here, we extend that
experiment into dense gas by probing the grain alignment around the
Becklin-Neugebauer Kleinmann-Low (BNKL) nebula in Orion using
polarization data at 100, 350, and $850\,\micron$.

\section{Analysis and Results} \label{sec-anal}

\begin{figure}
  \plotone{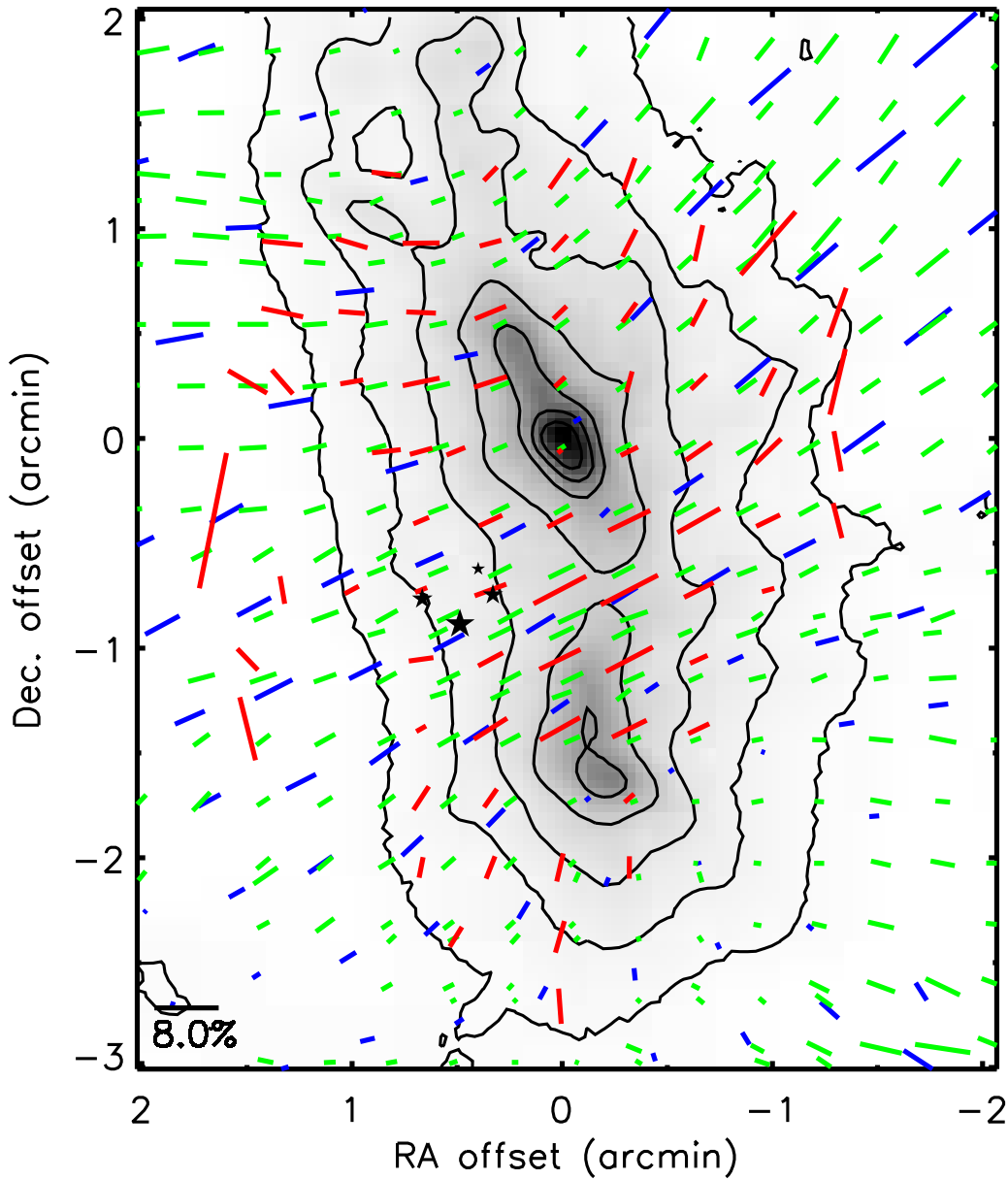}
  \caption{Polarimetry in the Orion Molecular Cloud. Colored lines
    show the $100\,\micron$ (blue), $350\,\micron$ (green), and
    $850\,\micron$ (red) polarization, drawn parallel to the
    inferred magnetic field orientation with lengths proportional to
    the polarization signal (lower-left scale bar).  For clarity
    only one-quarter of the $850\,\micron$ data are plotted.  The
    $350\,\micron$ total intensity, with $10\arcsec$ resolution
    \citep{vaillancourtetal2008}, is shown in grayscale with contours
    at 2, 5, 10, 20, 40, 60, and 80\% of the peak flux. Coordinate
    offsets are measured with respect to the peak $350\,\micron$
    intensity (IRc2). The Trapezium cluster
    is shown just east of the submillimeter intensity ridge.}
  \label{fig1}
\end{figure}

We use three polarization data sets (Figure~\ref{fig1}) at
wavelengths of $100\,\micron$ \citep{dotsonetal2000}, $350\,\micron$
\citep{dotsonetal2010}, and $850\,\micron$ \citep{matthewsetal2009}.
The radial distance to each data point and their spatial position
angles on the sky
are measured with respect to the position of peak submillimeter flux,
coincident with the source IRc2 ($\alpha = 5\hour35\minute14\fs5 $,
$\delta = -5\arcdeg22\arcmin29\arcsec$, J2000).  We limit the data to
points within 1.5 arcminutes of IRc2, a choice driven by the facts 1)
at larger distances south the Orion-S source ($\Delta\delta \sim
-1\farcm6$) may confuse the analysis and 2) smaller distances limit
the analysis to regions very close to BNKL, which may imprint
complicated density and turbulent effects on any signal.

The position angle of radiation from the BNKL region to any data point
is the position angle on the sky, measured east of north.  Even at
distances $<$$1\farcm5$, Orion-S may bias the analyses.  Therefore,
analyses do not include data within the 20\% $350\,\micron$ intensity
contour (Figure~\ref{fig1}) about Orion-S.

\subsection{Single Frequency Fits} \label{sec-single}

\begin{figure}
  \plotone{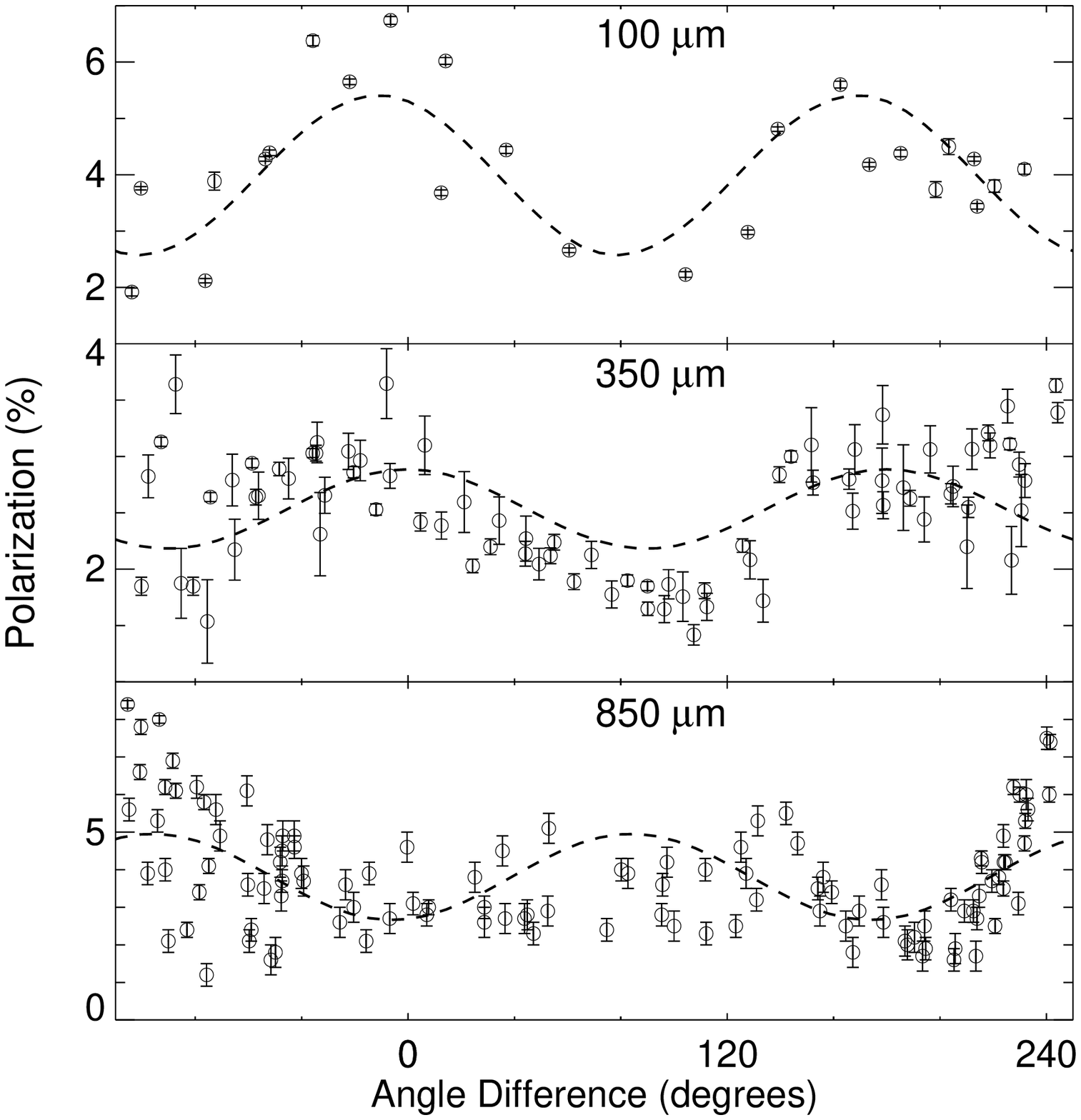}
  \caption{Polarization vs.\ angle. The polarization signals as a
    function of the angle difference between the magnetic field
    orientation and radiation direction ($\Psi\equiv\phi_\mathrm{rad}
    - \phi_\mathrm{B}$) are shown for the $100$, $350$, and
    $850\,\micron$ datasets. Data here are limited to radial
    distances from IRc2 in the range 0.5 to 1.5~arcminutes (see text
    for details).  Fits to $180\arcdeg$ sinusoids are shown as the
    dotted curves; fit parameters are listed in Table~\ref{tbl-1}.}
  \label{fig2}
\end{figure}

Figure~\ref{fig2} compares the fractional polarization to the angle
difference between the radiation position angle and the inferred
magnetic field orientation, $\Psi \equiv \phi_\mathrm{rad} - \phi_B$.
Fits of these data to sinusoids with a $180\arcdeg$ period provide a
test of the existence and likelihood of such signals.
Formal fits are of the form
\begin{equation}
p = p_0 + A\cos[2(\Psi-\delta)]
\label{eq-1}
\end{equation}
where $p_0$ is a constant offset, $A$ is the sinusoid amplitude, and
$\delta$ is the phase offset from the peak signal. Figure~\ref{fig2}
shows only data in the distance range $0\farcm5$--$1\farcm5$, while
Table~\ref{tbl-1} reports parameter values at other distances
(Section~\ref{sec-distance}). All fits use equally weighted data
points.

The sinusoidal signals in Figure~\ref{fig2} are fairly clear
``by-eye'' at 100 and $350\,\micron$, a finding confirmed by the
significance of the amplitude measurements ($A/\sigma_A>3$) and the
small phase uncertainties ($\sigma_\delta < 7\arcdeg$). The signal is
less clear at $850\,\micron$ and may be biased by large excursions near
$\Psi\sim260\arcdeg$,
although the amplitude
significance and phase uncertainty are similar to the other two
wavelengths. The fit phase angles are in good agreement at 100 and
$350\,\micron$, but orthogonal to that at $850\,\micron$.

\subsection{Periodograms} \label{sec-period}

The simple interpretation of a high significance $A/\sigma_A$ result
for the single-frequency fit is to confirm the presence of the model.
However, this is only true when compared against the
null-hypothesis---random noise with respect to a constant signal as a
function of $\Psi$.
One less restrictive subset of models 
are sinusoids at other single frequencies. We test fits to these other
frequencies using the generalized Lomb-Scargle periodogram for
unevenly spaced data as described by \citet{zechmeisterkurster2009}.
This analysis is equivalent to performing least-squares fits to every
frequency in turn \citep{scargle1989}.

A key parameter in the periodogram, and the subsequent probability estimate,
is the number of independent frequencies, $N_f$, in the data.  For
unevenly spaced data the number of independent frequencies can be on
order the number of data points, $N_d$
\citep{hornebaliunas1986}. Given the relative uncertainties and
``clumpiness'' of the data in $p$-$\Psi$ space this is a likely
upper limit so we choose $N_f=N_d$ (see Table~\ref{tbl-1}).
Since the data are limited to a real spatial circle any signal must
repeat at periods of $360\arcdeg$, so only periods with integer
fractions of $360\arcdeg$ may exist (360, 180, 120, ... $360/N_f$).

\begin{figure}
  \plotone{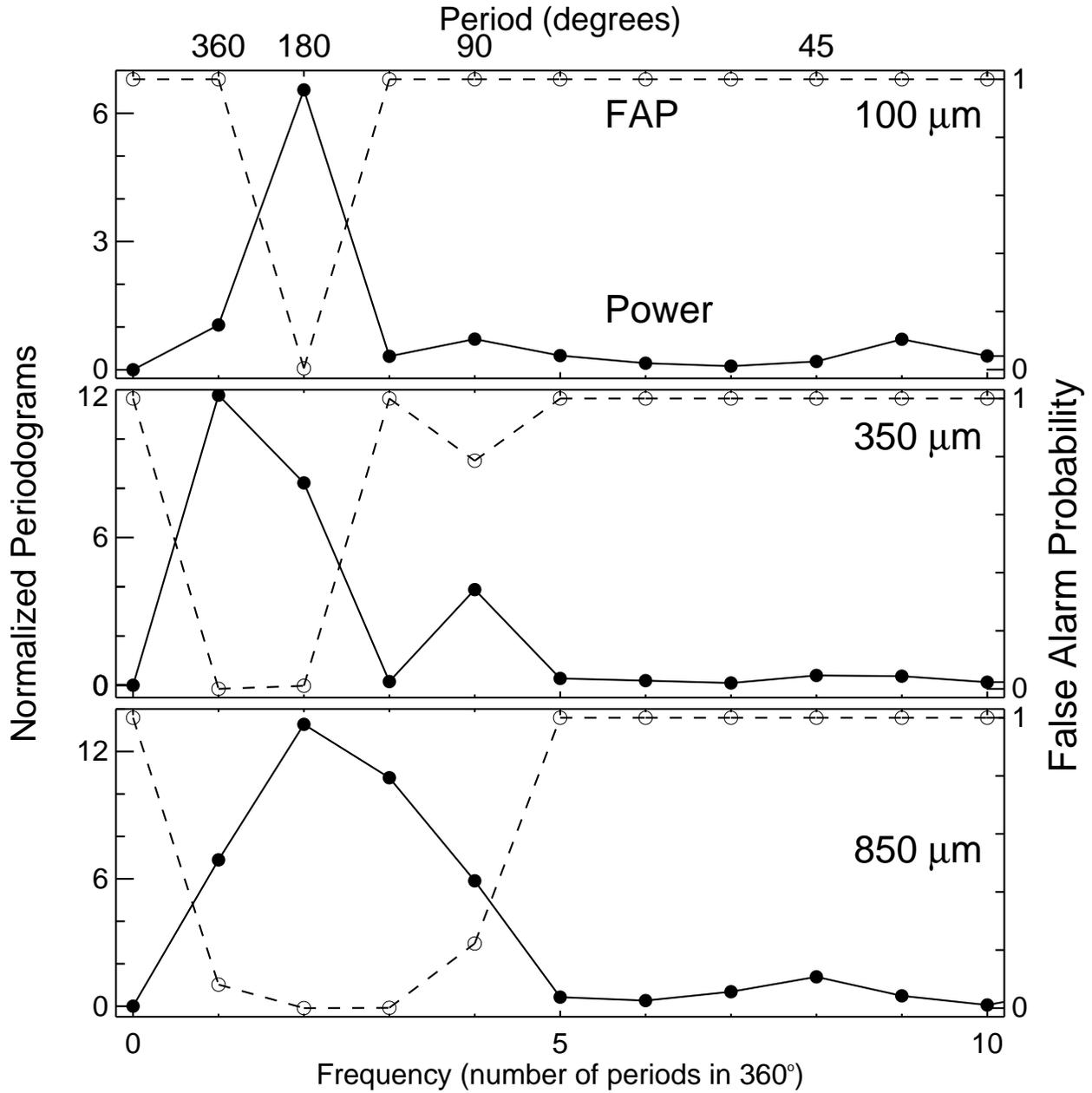}
  \caption{Periodograms and probabilities. The power spectra
    (solid lines, left axis) are shown as a function of frequency for
    the 100, 350, and $850\,\micron$ datasets of
    Figure~\ref{fig2}. The associated false alarm probabilities (FAP,
    dotted lines, right axis) are shown in the corresponding panels.}
  \label{fig3} 
\end{figure}

For radii $0\farcm5$--$1\farcm5$ from BNKL, the 100 and $850\,\micron$
periodograms peak at frequency $f=2$, a period of $180\arcdeg$
(Figure~\ref{fig3}). The $350\,\micron$ data peaks at $f=1$
($360\arcdeg$), although $f=2$ is the next largest power. The fact a
peak does not occur at $f=2$ does not imply that such a signal is
absent, but simply that it is not the dominant frequency in
the data.  The false alarm probability (FAP) measures whether
or not any given peak contributes significantly to the signal.

The FAP describes the probability that a signal would be observed in
the case of random noise and the absence of that signal.  The FAP
follows from $N_f$ and a properly normalized periodogram.  The powers
and FAPs in Figure~\ref{fig3} and Table~\ref{tbl-1} use the ``sample
variance'' normalization of \citeauthor{hornebaliunas1986}
(\citeyear{hornebaliunas1986}, also \citealt{zechmeisterkurster2009}).
We note that FAP increases for increasing $N_f$, thus the choice of
$N_f=N_d$ yields upper limits on the FAPs\@.  The smallest FAPs
correspond to the largest periodogram powers and confirm the presence
of a 180-degree component in the 100 and $850\,\micron$ data (FAPs =
0.4\% and 0.004\%, respectively).  However, frequency components away
from the peak can also have very small FAPs, indicating the presence
of multiple components.  For example, in the case of the
$350\,\micron$ data, the FAPs for the 360- and 180-degree components
are 0.01 and 1.0 percent, respectively. The $360\arcdeg$
component is likely the result of localized spatial features overlayed
on the periodic signal, thereby breaking the $180\arcdeg$ symmetry and
moving power to other frequencies.




\section{Discussion} \label{sec-3}

\subsection{Comparison to Predictions} \label{sec-disc1}

\subsubsection{The 180 degree Component}

The radiative torque model predicts an increased alignment efficiency
when the direction of the radiation, $\phi_\mathrm{rad}$, is parallel
to the grain spin axis and to the magnetic field orientation, $\phi_B$
\citep{hoanglazarian2009b}.  
In the absence of changes in collisional disalignment and magnetic
field orientation 
the measured polarization fraction is a measure of this
alignment efficiency.
Thus one expects the variation in polarization as a function of $\Psi
= \phi_\mathrm{rad}-\phi_B$ to exhibit a signal with a $180\arcdeg$ period.


The $100\,\micron$ periodogram has a strong peak at the expected
$180\arcdeg$ period, has a low FAP ($<$2.6\%), and the phase angle
$\delta$ peaks close to $\phi_\mathrm{rad} = \phi_B$
(Table~\ref{tbl-1}). While the $350\,\micron$ periodogram peaks at a
$360\arcdeg$ period, the $180\arcdeg$ component also has a low FAP
($<$3.1\%). Additionally, the $350\,\micron$ phase angle shows good
agreement between $\phi_\mathrm{rad}$ and $\phi_B$
($\sim$$0\arcdeg$--$4\arcdeg$ with uncertainties
$4\arcdeg$--$7\arcdeg$).
The $850\,\micron$ phase angles are similar in the two lower radius
bins ($0\farcm0$--$1\farcm5$ and
$0\farcm5$--$1\farcm5$) but are orthogonal to both
the expected RAT signal and the results at 100 and $350\,\micron$.  
As the emission is optically thin ($\tau[100\,\micron]<0.4$;
\citealt{cole1999}), the three wavelengths sample
dust along the entire line-of-sight.  However, they sample
different environments since the shorter wavelengths are more
sensitive to hotter dust closer to the photon sources. The heating of
the warmer/closer dust is dominated by BNKL while heating of the
cooler dust likely has contributions from diffuse radiation outside
that immediate environment.

While the use of a sinusoid to characterize the variation is chosen
for simplicity, any symmetric 180-degree signal (e.g., a triangle or
square wave) must contain this fundamental frequency.
The limitation in searching for a single frequency is a loss in
observed power, making it less likely to observe a signal in its
absence. The periodogram analysis provides a natural way to find this
missing power.

The analysis presented here can also be applied to the
optical polarization data around the star HD\,97300, where
\citet{Anderssonetal2011} found a strong signal and peak alignment
efficiency within $10\arcdeg$ of $\phi_\mathrm{rad}=\phi_B$.
Similarly, we find that the $180\arcdeg$ component dominates the
HD\,97300 data with $\mathrm{FAP} = 6\%$ and phase $\delta =
10\arcdeg\pm7\arcdeg$.

\subsubsection{Other Frequency Components}

The single-frequency fits assume the $180\arcdeg$ signal is the
only signal present. However, 
at $350\,\micron$ the $360\arcdeg$ and $180\arcdeg$ components have
FAPs of $<$1.0\%, and at $850\,\micron$ the $180\arcdeg$ and $120\arcdeg$
components have FAPs $<$0.1\%. This indicates the simultaneous
presence of multiple components such that one component cannot
necessarily be ignored when fitting for the other. Formally, including
a $360\arcdeg$ component to the $350\,\micron$ data and including a
$120\arcdeg$ component to the $850\,\micron$ do significantly improve
the fits (as measured by an $F$-test). However, the corresponding
$180\arcdeg$ amplitudes and phases vary only within the
uncertainties reported in Table~\ref{tbl-1}. Therefore, the presence
of multiple frequency components can neither change any conclusions
with respect to $\Psi$, nor can it explain the large phase offset for
the $850\,\micron$ data.

The physical interpretation of a $360\arcdeg$ component is simply the
fact that data
are not equivalent when reflected about the origin.
For example, the OMC-1 intensity map is clearly asymmetric in the
north-south direction, with Orion-S dominating in the south.  If this
radiation source has an influence on grain alignment and/or
polarization, then one can expect some significant signal at
$360\arcdeg$ and for that signal to get stronger as data closer to
Orion-S are included.
However, the limited data (three radius bins) are insufficient to test
this hypothesis.

We have attempted to limit the possible effects from Orion-S by removing data in a
small region around it (Section~\ref{sec-anal}).  However, some
$850\,\micron$ data in that direction show large excursions from the
best single-frequency fit at angles $\Psi\lesssim -70\arcdeg$ and
$\Psi \gtrsim 220\arcdeg$ (Figure~\ref{fig2}).  These data are mostly
located in spatial regions towards Orion-S with $\phi_\mathrm{rad}
\approx 130\arcdeg$ -- $235\arcdeg$ and distances beyond $0\farcm5$.
Removing those data from the analysis (at radii
$0\farcm5$--$1\farcm5$) removes the $360\arcdeg$ and $120\arcdeg$
components from the $850\,\micron$ data (FAPs $\sim$99\% at both
$360\arcdeg$ and $120\arcdeg$).  At $350\,\micron$ the both the
$360\arcdeg$ and $180\arcdeg$ components remain and the $180\arcdeg$
phase angle ($\delta=-12\arcdeg\pm3\arcdeg$) is in better agreement
with that at $100\,\micron$.  This change has little effect on the
conclusion that the $180\arcdeg$ component exists with high confidence
at $850\,\micron$.  Changes in the $100\,\micron$ data fits are minimal,
with values consistent with those in Table~\ref{tbl-1}, but the FAP is
increased to 12\%.



\subsection{Distance Dependence} \label{sec-distance}

If grain alignment is due to radiative torques, and no other factors
contribute significantly to either alignment or disalignment, then the
alignment decreases with distance from the radiation source. We divide
each dataset into three distance bins with respect to BNKL, all with
upper limits of $1\farcm5$ but varying lower limits of $0\farcm0$,
$0\farcm5$, and $1\farcm0$ and repeat the analyses in
Section~\ref{sec-anal} (Table~\ref{tbl-1}). FAPs
are low ($<$7.5\%) for all wavelengths and radii. The shorter
wavelengths see an increase in the amplitude and drop in FAP in the
largest distance bin only. The $850\,\micron$ amplitudes increase
across all three bins, although the largest has a relatively high FAP
(7.5\%).

We attribute the increasing signal, as opposed to the expected
decrease, to lower polarization values towards the core regions of
BNKL\@. This drop in polarization is either the result of loss of
grain alignment due to the large collision rate at higher space
densities or large changes in polarization position angle, either along
the line of sight or within the telescope beam.

\begin{figure*}
  \plotone{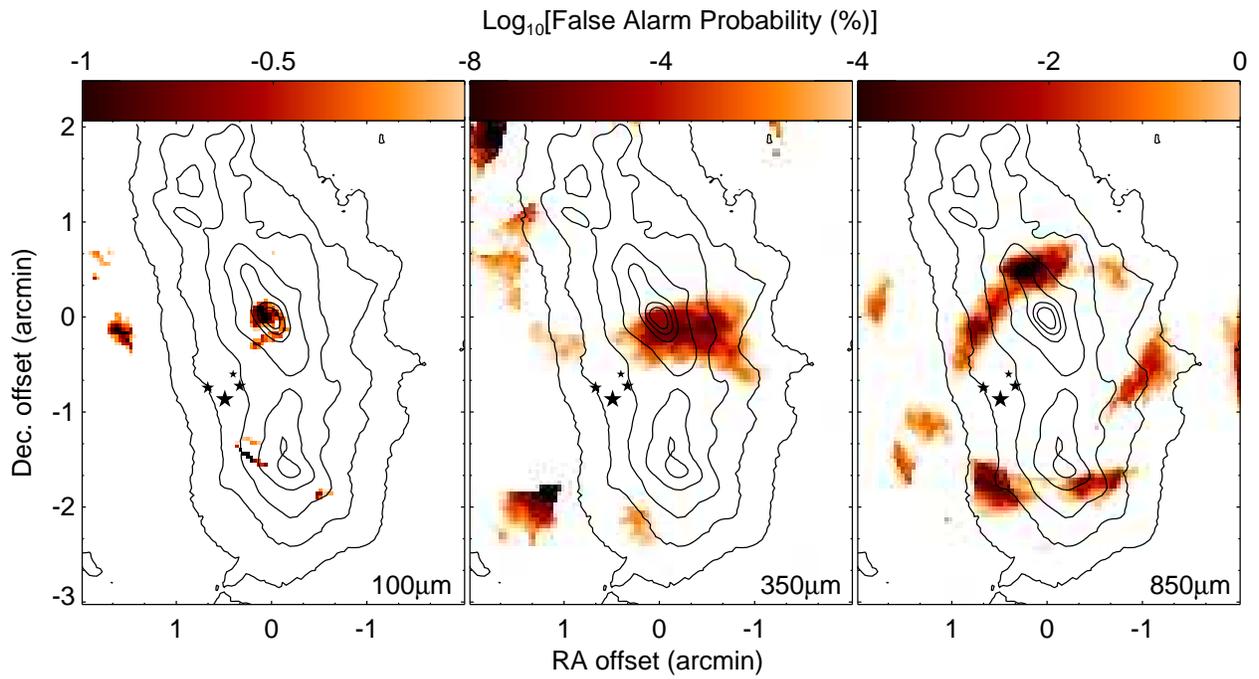}
  \caption{False alarm probabilities (FAP) for all OMC-1 data for the
    $180\arcdeg$ component at radii $1\farcm0$--$1\farcm5$.  The
    logarithmic grayscales (at top) show the FAPs with a maximum value
    of 1\% in all panels ($\log[1\%]=0$). Coordinate offsets, Trapezium stars, and
    $350\,\micron$ intensity contours are as in Figure~\ref{fig1}.}
  \label{fig4}
\end{figure*}


\subsection{A Systematic Search}

Section~\ref{sec-anal} measures the directional dependence
of RAT centered only on the submillimeter intensity peak BNKL/IRc2.
It is reasonable to ask whether there are other such sources, and
whether a signal would be obtained in other regions of the map.

Figure~\ref{fig4} shows the FAPs across the entire OMC-1 map,
generated by applying the same periodogram analysis to every point in
the map.  Here we show the analysis for the $180\arcdeg$ component
within the $1\farcm0$--$1\farcm5$ radius bin at each point. This
choice provides qualitatively similar results to those at
$0\farcm5$--$1\farcm5$ but with an increased contrast for clarity.
Note that the data set itself covers a region larger than that shown
in Figure~\ref{fig1}, so FAPs at the apparent edge of
the map are still valid with this annulus.

The FAPs are relatively high at all points in the map except for a few
scattered regions. It is reasonable to expect that, by chance, some
significant signal would be observed, especially in low intensity
regions where uncertainties are larger. The proximity of the low FAPs
to BNKL lends further support to the analysis already presented. At 100
and $350\,\micron$ there are clear signals towards BNKL, as expected
for the RAT hypothesis. Possible signals also exist in an extended
region at $350\,\micron$ and just to the east of Orion-S at
$100\,\micron$.

The $850\,\micron$ FAP towards BNKL is very high (20--60\%) but low in
areas just outside of BNKL and Orion-S.  This is consistent with the earlier hypothesis
(Section~\ref{sec-disc1}) that cooler dust emitting at the longer
wavelengths and further from the strong FIR sources has strong heating
and RAT contributions from sources other than BNKL.

There are no strong signals towards the Trapezium stars, which might
be expected to generate high RATs due to photons from the bright OB
stars. Since those stars are further from the dust ridge (0.25 --
0.5\,pc; \citealt{odell2001}) than BNKL, and the optical light is heavily extincted by that
dust, they have little influence on the heating of grains observed in
the FIR.

\subsection{Angle Projection} \label{sec-3d}

The input to the RAT model is the three-dimensional space angle
$\Psi_3$ between the magnetic- and radiation-field directions.  In
this work we measure the two-dimensional projection onto the plane of
the sky, $\Psi_2$ (previously defined as $\Psi$).
From the dot product of the
radiation and magnetic field space vectors we have
\begin{equation}
\cos \Psi_3 = \cos\Psi_2 \sin\theta_\mathrm{rad} \sin\theta_B  +
\cos\theta_\mathrm{rad} \cos\theta_B,
\label{eq-angle}
\end{equation}
where $\theta_{\mathrm{rad},B}$ are the inclination angles of the
field vectors with respect to the line of sight. The largest signal
with respect to $\Psi_2$ occurs when both vectors are in the plane of
the sky ($\theta_\mathrm{rad} = \theta_B = 90\arcdeg$) but becomes
unmeasurably low when one vector lies near the line of sight
($\theta=0$). For our work here it is important to note $\Psi_2$
rotates with the same period as $\Psi_3$, just with different
amplitudes.

The projection effect is fairly small since the magnetic field in OMC-1 has an
inclination angle of $\approx$$65\arcdeg$
\citep{houdeetal2004a}. However, the projected radiation angle may be
a stronger effect.  
For the line-of-sight distance estimate we assume the cloud depth is
approximately the same as its width and use a half-width $3\farcm5$
\citep{houdeetal2009}.
Thus, the inclination angle is as low as
$\phi_\mathrm{rad}=\arcsin([0.5$--$1.5]/3.5)=8\arcdeg$--$25\arcdeg$ at the edge
of the cloud.  However, integration through the cloud depth
likely shifts the average angle to larger values since the angle
increases for dust closer to the cloud center and the central regions
dominate the integrated column density.

The effects of line-of-sight averaging and binning in the plane of the
sky lowers any signal as the dispersion increases within any given
bin. However, if this had a large effect on these data, no
azimuthal signal would be detected. The existence of strong signals,
and at multiple wavelengths, implies such averaging and projection
effects do not strongly effect our conclusions.

\section{Summary} \label{sec-summary}

In radiative grain alignment anisotropic radiation provides the
torques to align the grains with interstellar magnetic fields. The
model makes several observable predictions, including a more efficient
alignment if the direction of the radiation field anisotropy,
$\phi_\mathrm{rad}$, is parallel to the magnetic field orientation,
$\phi_B$. In the case of $\lambda=100\,\micron$ data about the bright
source BNKL/IRc2 in Orion, the polarization fraction peaks within
$15\arcdeg$ of these angles. For $350\,\micron$ data the agreement is
also good at $<$$4\arcdeg$, but the two directions are orthogonal at
$850\,\micron$.

Periodogram analyses show that the polarization fraction as a function
of the angle difference $\phi_\mathrm{rad}-\phi_B$ at all three
wavelengths is dominated by a signal with a period of $180\arcdeg$, as
expected for a RAT. In most cases, the probabilities that these
signals arise from random noise (the false alarm probabilities, FAPs)
are less than 1\%.
Contributions from signals with other periods do not significantly
change the quantitative values of the fit amplitudes, phases, or FAPs.
The map of the FAPs has a narrow spatial minimum towards BNKL at
$100\,\micron$, is more extended at $350\,\micron$, but is outside the
core at $850\,\micron$. This drop in signal with increasing wavelength
is expected since grains emitting predominantly at $100\,\micron$ are
warmest, and thus closest, to the BNKL source. Emission at longer
wavelengths arises from cooler grains further from the source
with increased heating contributions from sources outside BNKL\@.

Based on the low FAPs for $180\arcdeg$ periods and the good agreement
between angular peaks at 100 and $350\,\micron$ we find strong
evidence for a detection of the predicted angular dependence of RATs 
This work, taken together with the optical work towards HD\,97300
\citep{Anderssonetal2011}, provides the best current tests of the
RAT-predicted angular alignment dependence.




\acknowledgments

This work has been supported by NSF grant AST 11-09469.


\begin{deluxetable}{cccccccccc}
\tabletypesize{\scriptsize}
\tablecaption{Model Fit Parameters\label{tbl-1}}
\tablewidth{0pt}
\tablehead{
  \colhead{Wavelength} & \colhead{Radii} & 
  \colhead{$p_0$} & \colhead{$\sigma_{p_0}$} &
  \colhead{$A$} & \colhead{$\sigma_A$} & 
  \colhead{$\delta$} & \colhead{$\sigma_\delta$}
  & \colhead{FAP\tablenotemark{a}} & \colhead{$N_d$\tablenotemark{b}} \\
  \colhead{($\micron$)} & \colhead{(arcmin)} & 
  \colhead{(percent)} & \colhead{(percent)} &
  \colhead{(percent)} & \colhead{(percent)} & 
  \colhead{(degrees)} & \colhead{(degrees)} &
  \colhead{(percent)}
}
\startdata
100 & 0.0 -- 1.5 & 3.8 & 0.2 & 1.4 & 0.3 & $-15$    & 6 & 2.6\phn & \phn27\\
    & 0.5 -- 1.5 & 4.0 & 0.2 & 1.4 & 0.3 & $-11$    & 5 & 0.4\phn & \phn25\\
    & 1.0 -- 1.5 & 4.0 & 0.2 & 2.1 & 0.3 & $-10$    & 4 & 0.08    & \phn14
    \\ \\
350 & 0.0 -- 1.5 & 2.5 & 0.1 & 0.3 & 0.1 & \phn$-4$  & 7 & 3.1\phn & \phn91\\
    & 0.5 -- 1.5 & 2.5 & 0.1 & 0.3 & 0.1 & \phn\phs0 & 7 & 1.0\phn & \phn82\\
    & 1.0 -- 1.5 & 2.4 & 0.1 & 0.6 & 0.1 & \phn$-2$  & 4 &
    $10^{-5}$ & \phn51\\ \\
850 & 0.0 -- 1.5 & 3.6 & 0.1 & 0.9 & 0.2 & \phs86   & 6 & 0.2\phn\phn & 136\\
    & 0.5 -- 1.5 & 3.8 & 0.1 & 1.1 & 0.2 & \phs84   & 5 & 0.004       & 115\\
    & 1.0 -- 1.5 & 3.0 & 0.2 & 1.3 & 0.3 & \phs69   & 7 & 7.5\phn\phn & \phn41
\enddata
\tablecomments{Parameters and uncertainties fit to the
  180-degree period signal with offset $p_0$, amplitude $A$, and phase $\delta$
  (equation~[\ref{eq-1}]). Fits include only data within the
  noted radial distances from BNKL\@. The $A$ and
  $\sigma_\delta$ values have been corrected for statistical bias 
  \citep{vaillancourt2006}.}
\tablenotetext{a}{The False Alarm Probability (FAP) is the probability
  a signal would be observed in the case of random noise.}
\tablenotetext{b}{Number of data points used in analysis.}
\end{deluxetable}

%


\end{document}